\documentclass[12pt]{iopart}
\usepackage{graphicx}
\usepackage{epstopdf}
\usepackage{color}
\usepackage{epsfig}
\usepackage{cite}
\usepackage{amssymb}

\newcommand{\bea}{\begin{eqnarray}}
\newcommand{\eea}{\end{eqnarray}}
\newcommand{\ket}[1]{\left|{#1}\right\rangle}
\newcommand{\bra}[1]{\left\langle{#1}\right|}
\newcommand{\aver}[1]{\left\langle{#1}\right\rangle}

\usepackage{caption} 
\usepackage{enumerate}
\usepackage{float}

\begin{document}
\title{Dynamics of an open quantum system interacting with a quantum 
environment}
\author{Athreya Shankar, S Lakshmibala and V Balakrishnan}
\address{Department of Physics, Indian Institute of Technology Madras, 
Chennai 600 036, India}
\begin{abstract}
We examine the dynamics of subsystems of bipartite and tripartite quantum 
systems with nonlinear Hamiltonians.  
We consider two models which capture the generic features of open 
quantum systems:     
a three-level atom interacting with a single-mode radiation field, 
and a  
three-level atom interacting with two field modes which do not directly interact 
with each other. The entanglement of specific initially unentangled 
states of the atom-field system is examined 
through the time-varying subsystem von Neumann entropy (SVNE). The counterparts of  
near-revivals 
and fractional revivals of the initial state are clearly identifiable in the SVNE in 
all 
cases where revival phenomena occur. The Mandel $Q$ parameter 
corresponding to the photon number of a radiation field is 
obtained as a function of time  in both models. In those cases where 
revivals are absent, a  time series analysis of   
the mean photon number reveals a variety of ergodicity 
properties (as manifested in return maps, recurrence-time distributions 
and Lyapunov exponents), depending on the strength of the 
nonlinearity and the degree of coherence 
of the initial state of the radiation field(s).   
\end{abstract} 
\maketitle
\section{Introduction}
 \paragraph{} Quantum systems governed by nonlinear Hamiltonians can display a 
wide 
variety 
of 
nonclassical effects such as revivals and fractional revivals during their 
temporal evolution \cite{robi}. The dynamics  of subsystems of quantum systems 
 depends strongly 
on the nature of their interactions with the quantum 
`environment' to which they are exposed. The crucial point is the following.
The full system remains in a pure state whose unitary time development is  
governed by an 
appropriate hermitian Hamiltonian. A subsystem, however, is 
described, in general, by 
a reduced density matrix that evolves through a dynamical map.
This, in turn, could result in lossy interaction of the subsystem with the 
environment, 
leading to decoherence effects. The subsystem can effectively be modelled as a 
dissipative system.

 A convenient framework to examine the rich dynamics of an open 
quantum subsystem is 
provided by specific bipartite and multipartite  models of atom-field interactions.
As we shall see, the dynamics of the field in  such models 
differs considerably  
from that of an initial single-mode coherent state (CS) of the radiation field   
propagating in a nonlinear (Kerr) medium 
with an effective Hamiltonian
 \cite{milburn,kita}
\begin{equation}
H_{\rm Kerr} =\hbar \chi a^{\dag ^2}{a}^2.
\label{kerrham}
\end{equation}
Here, $a$ and $a^\dagger$ are the usual photon annihilation and 
creation operators, and $\chi$ is a positive constant that sets a time scale. 
Exact revivals of the field state occur  in this case at instants of time that are 
integer multiples of $\pi / \chi$: at these instants an  initial state returns to 
itself apart from an overall phase. 
Hence all expectation values also return to their 
initial values. Under specific conditions, fractional revivals could occur between 
two successive revivals: at these instants, the  
wave packet splits into a number of spatially distributed subpackets each of which 
closely resembles the initial wave packet. 
Revivals have been understood in a general setting \cite{aver1} 
for an arbitrary 
initial superposition of  photon number  states,  
and  revival phenomena have been discussed in a wide class of
systems \cite{robi,milburn,kita}, \cite{greiner,tara,tanas,yurke}.   
In most of these earlier studies    
the initial state is a 
CS $\ket{\alpha}$, where $a \ket{\alpha} = \alpha \ket{\alpha}$
and $\alpha\in\mathcal{C}$.

It has further been established \cite{sudh1, sudh2} that in the case of the  
Kerr Hamiltonian in (\ref{kerrham}), 
signatures of the 
occurrence of revivals
and
fractional revivals of the radiation field
are manifested in the
mean and higher moments of appropriate observables.
The field states considered in this case are CS 
as well as  
photon-added coherent states (PACS). 
The $m$-photon-added coherent state ($m$-PACS) $\ket{\alpha, m}$ 
is 
obtained 
by an $m$-fold  application 
of the photon creation operator on the CS $\ket{\alpha}$,  
and normalizing the resultant state \cite{tara}. 
The PACS family 
displays 
precisely quantifiable 
departure from coherence, and hence lends itself to a systematic examination of the 
role 
of coherence in wave
packet dynamics. Experimental identification and 
characterization of the single 
photon-added coherent state \cite{zavatta} using quantum state tomography has added 
more impetus to such studies. 

In 
contrast 
to this, even in the simplest bipartite atom-field system  
modelled by a 
nonlinear Hamiltonian involving both field and atom operators explicitly, 
exact revivals of the initial state need not occur. The dynamics in this case is 
enriched by the phenomenon of quantum entanglement. Even if the initial state
of the system is an unentangled direct product of the field and atom states, 
entanglement occurs during 
temporal evolution, and exact revivals of the initial state are generically absent. 
Wave packet revivals  would be even less probable in a multipartite system, 
in general. 

The occurrence of near-revivals therefore is the best that one can hope for in 
general in these cases. In bipartite models of atom-field interactions, the 
appearance of such near-revivals would be crucially 
dependent on the ratio of the 
strength of the nonlinearity to the strength of the interaction between the two 
subsystems. A  model Hamiltonian describing a multi-level  
nonlinear atomic medium interacting with a single-mode  radiation field \cite{puri}
has been examined earlier \cite{sudhjphysb}, illustrating 
when near-revivals 
occur in bipartite 
systems, and how 
the subsystem von Neumann entropy (SVNE) 
and the system linear entropy (SLE) mirror the appearance of revivals and 
fractional revivals. These results indicate that, for an initial coherent state 
of the field,   
 if the strength of the nonlinearity is significantly smaller 
than the strength of atom-field interaction,  both the SVNE and SLE show pronounced 
dips 
at the near-revival time period $T_{rev}$, and 
fractions 1/2, 
1/3, 1/4 and 2/3 of $T_{rev}$ when fractional revivals occur. In contrast, for the 
same numerical values of the nonlinearity and coupling strengths, if the initial 
state of 
the radiation field is a PACS, the SVNE at any instant also increases in comparison 
with 
the case of an initial CS, and even near-revivals disappear. 

The latter situation raises interesting questions 
pertaining to  
the dynamical behaviour  of the expectation values of observables. For instance, an 
experimentally relevant quantity
such as the mean photon number of a radiation 
field \cite{webb} could move far away 
from its 
initial value over a sufficiently long period of time, in a situation where even 
near-revivals are absent. A time series analysis of the mean photon number would 
reveal a diversity of ergodicity properties of the observable. Such an 
investigation
has been carried out \cite{sudhergodicity} in the framework of the 
bipartite model considered in \cite{puri}. It is found that, 
if the 
strength of the nonlinearity is significantly more than the interaction 
strength, then,  depending on the level of departure from coherence of the initial 
state, the mean photon number could even display exponential instability 
indicated by a 
positive Lyapunov exponent obtained through the time series analysis.

Extensive work has been carried out to understand the ergodicity properties of 
classical dynamical systems, by examining  return-time statistics  of  
dynamical variables to coarse-grained cells in phase space, and rigorous 
results established on 
Poincar\'{e} recurrences \cite{kac}. 
Important and interesting results are known on the recurrence properties of
classical
systems such as
Hamiltonian systems,
measure-preserving maps, and  dissipative maps. 
In conservative classical systems
it has been
established that universal asymptotic properties 
including power-law recurrence-time
distributions arise due to the non-uniform nature of invariant sets in phase space
and `stickiness' of remnants of invariant tori \cite{chirikov}.  Again, detailed  
studies of recurrence-time distributions in low-dimensional maps enable us to    
differentiate clearly between varying degrees of randomness ranging from 
quasiperiodicity through intermittent behavior to fully-developed chaos 
\cite{nicolis, nicolis1}. 
Apart from these,
recurrence plots \cite{eckmann} are used to analyze
the dynamical behaviour of classical variables.

Similar in-depth investigations on the behaviour of quantum observables   
are scant. 
As is the case in classical systems,  we would expect recurrence time  
statistics of quantum observables to complement information obtained through 
Lyapunov 
exponents deduced from a time series analysis of the observables.
To carry out such studies, it is appropriate to  consider the recurrence 
time distributions
of the expectation values of suitable  observables (such as the mean photon 
number)  in models of 
atom-field interactions,  to cells in a  `phase
space' of these expectation values. 
An inherently quantum 
mechanical feature that arises in this approach 
is that the phase space 
is now effectively infinite-dimensional,  as it involves the expectation values of  
all the relevant observables,  
their higher moments, and all correlators. 
In principle, the collective dynamics of all these variables  
needs 
to be analyzed.
In practice, 
therefore, it is crucial to identify an 
adequately tractable and experimentally relevant observable (or a minimal set 
of such observables) whose dynamical properties may  be investigated. 
When a quantum mechanical system is partitioned into subsystems, each subsystem can 
interact with the others (which constitute a quantum mechanical `environment') in a 
complicated manner.
The dynamics of observables will depend quite strongly on the specific initial 
state. The ergodicity properties of these observables can be quite complex. Such an 
approach facilitates the understanding of the ergodic behaviour of classical and 
quantum systems in a unified manner, and enables us to   
relate known results
on Poincar\'{e} recurrences obtained from classical ergodic theory, on the one hand,
and the temporal behaviour of quantum expectation values  treated as `classical' 
dynamical
variables, on the other.
     
The results of such an investigation on the nonlinear bipartite model  
 \cite{puri} of a multi-level atom (modelled by a nonlinear oscillator) interacting 
with a single-mode field in an initial CS or 
a PACS  
indicate \cite{sudhergodicity,sudhrecurrence} that the first-return-time 
distribution 
is spiky for 
merely quasiperiodic dynamics,  and exponential for  
long-term chaotic behaviour of the mean photon number.
In this bipartite model, the Hilbert spaces of 
both 
the radiation field and the atom are essentially infinite-dimensional. 
The behaviour of the SVNE and a time series analysis 
of the mean photon number may be expected to  be significantly different from the 
foregoing in the case of a more realistic three-level atom (implying an 
associated  finite-dimensional 
Hilbert space) interacting with a radiation field. \`A priori we would 
expect that, for essentially the same ratio of the strength of the nonlinearity to 
that of the 
interaction,  marked differences in the dynamics of the mean photon number 
would occur, as compared to those reported in \cite{sudhergodicity}. 

 Going further, we may  generalize this bipartite system to one in which a 
three-level
$\Lambda$- or V-type atom interacts with two radiation fields which do not directly 
interact with each other.
Naturally, this tripartite system  may  be expected to display  far richer 
dynamics than the bipartite models. While both the field modes  have associated 
Hilbert 
spaces which are infinite-dimensional, the interaction proceeds through the very 
small 
number of 
channels supplied by the atom. This `bottleneck' may also be expected to affect 
the dynamics strongly. We are also motivated by the fact that 
photon-counting experiments 
on systems comprising a single atom interacting with laser light  
have already been realised (see, for instance, \cite{mckeever}.

In this paper, we examine the dynamics of the field mode in  two 
models describing the interaction between light  
 and a three-level V-type atom. The first 
is a 
bipartite model of a single-mode radiation field interacting with the  
atom so as to enable transitions from either of the excited levels to the ground 
state of the atom. The second is a tripartite extension of this model, 
where two independent single-mode fields interact with the $V$-type atom. 
Denoting the excited states of the atom by $\ket{1}$ and $\ket{2}$, and the 
ground state by $\ket{3}$, one of the radiation modes induces 
$\ket{1}  \leftrightarrow \ket{3}$ transitions, and the other induces $\ket{2} 
\leftrightarrow 
\ket{3}$ transitions. 
The ratio of 
the strength of
the
nonlinearity to the strength of the atom-field interaction is a controlling 
parameter.
In both models the Hamiltonian is  nonlinear in the field operators.   The initial 
states of the field(s)  that we consider are  CS and PACS. 

 In Section 2, we examine the SVNE as a function of 
time for initially unentangled atom-field states, and obtain results on the 
subsequent entanglement dynamics as well as near-revival phenomena. We also 
find the time variation of the 
Mandel $Q$ parameter (a measure of the deviation of the photon number from Poisson 
statistics). 
Wherever applicable, a 
comparison is made with corresponding results reported in \cite{sudhergodicity} on 
the dynamics of a multi-level atom interacting with a single-mode radiation field. 

In Section 3, we investigate the ergodicity properties of the mean photon number of 
the field in both these models for several different initial states. These show 
diverse kinds of behaviour, ranging from regular 
to chaotic, in the dynamics of the mean photon number of either field in the 
tripartite system. For our purposes, we have examined the mean photon number of the 
mode that induces  $\ket{1} \leftrightarrow \ket{3}$ transitions in the atom.   

In all the cases we consider,  the strength of the nonlinearity has been set equal 
to a value which is significantly 
higher
than the corresponding atom-field coupling  strength. 
As in the case of \cite{sudhergodicity}, it turns out that such a parameter ratio  
regime ensures non-trivial dynamical behaviour of the mean photon number such as 
exponential instability, near-revivals, and so on. 
We conclude with a summary of the results and comment on  possible
experimental verification of some of the salient features deduced.

\section{Entanglement dynamics in the bipartite and tripartite models}
\paragraph{}
We begin with the 
bipartite model describing the interaction of a single-mode radiation field 
with a three-level V-type atom. Transitions occur between  the two excited 
states and the ground 
state. Subsequently,  we extend the model to a tripartite 
Hamiltonian where two different field modes couple to the two different  
excited states and induce transitions between these and the ground state. In both 
cases the 
field 
operators exhibit  Kerr-type nonlinearity. 
Both the models  that we consider  
are simplified versions of models  which incorporate intensity-dependent couplings 
and 
non-zero detuning. For 
our present purposes, however, it suffices to take the interaction strength to be 
independent 
of 
the intensity of the field and set the detuning parameter to zero.
The three-level V-type atom consists of a ground state, labelled $\ket{3}$, and 
two
excited states $\ket{1}$ and $\ket{2}$. Only $\ket{3} \leftrightarrow \ket{1}$,
and the $\ket{3} \leftrightarrow \ket{2}$ transitions are allowed. Direct
$\ket{1} \leftrightarrow \ket{2}$ transitions between the excited states are 
forbidden.

\subsection{Interaction with a single field mode}

The general Hamiltonian for the 
bipartite system which includes an intensity-dependent coupling and non-zero 
detuning is given (see, e.g.,  \cite{zait2003nonclassical}) by 
\begin{eqnarray}
\fl H = \sum\limits_{j=1}^{3}\omega_{j}\sigma_{jj}
+  \Omega a^{\dagger}a  +  \chi a^{\dagger2}a^{2}
+  \lambda_{1}(R\sigma_{13} + R^{\dagger}\sigma_{31}) \nonumber \\
+  \lambda_{2}(R\sigma_{23} + R^{\dagger}\sigma_{32}).
\label{eqn:single_mode_v_hamiltonian}
\end{eqnarray}
Here, $\sigma_{jj} = \ket{j}\bra{j}$  where $\ket{j}$ is an atomic state,     
$\{omega_{j}\}$ are positive constants, 
$\Omega$ is the frequency of the
field mode, $\chi$ is the anharmonicity parameter, and $\lambda_{1}$ and   
$\lambda_{2}$ are the field-atom coupling strengths 
controlling
$\ket{3} \leftrightarrow \ket{1}$ and $\ket{3} \leftrightarrow \ket{2}$
transitions, respectively. $a^{\dagger}$ and $a$ are the photon creation 
and 
annihilation  operators, $N =
a^{\dagger}a$, and
$R = af(N)$, where 
$f(N)$ is a real-valued function that characterizes  
the intensity-dependent coupling. We set $\hbar = 1$ 
throughout. The Hamiltonian $H$  in \eref{eqn:single_mode_v_hamiltonian} is 
split
into two mutually commuting parts according to  $H = H_{0} + H_{1}$, 
where
\numparts
\begin{eqnarray}
H_{0} = \omega_{3} I  +  \Omega N^{\rm tot},
\label{eqn:single_mode_v_h0}
\end{eqnarray}
\begin{eqnarray}
\fl H_{1} = \chi  a^{\dagger2} a^{2}  -  \Delta_{1} \sigma_{11}
-  \Delta_{2} \sigma_{22}  +  \lambda_{1}(R\sigma_{13}
+ R^{\dagger}\sigma_{31}) \nonumber \\
+  \lambda_{2}(R\sigma_{23} + R^{\dagger}\sigma_{32}).
\label{eqn:single_mode_v_h1}
\end{eqnarray}
\endnumparts
Here $I = \sum_{j = 1}^{3} \sigma_{jj}$ 
and $N^{\rm tot} = a^{\dagger}a + \ket{1}\bra{1} + \ket{2}\bra{2}$ 
(in a sense,
 the `total number operator'). 
$\Delta_{1}$ and $\Delta_{2}$ are detuning
parameters given by
\numparts
\begin{equation}
\Delta_{1} = \omega_{3} - \omega_{1} + \Omega
\label{eqn:single_mode_v_delta1}
\end{equation}
\begin{equation}
\Delta_{2} = \omega_{3} - \omega_{2} + \Omega.
\label{eqn:single_mode_v_delta2}
\end{equation}
\endnumparts
As already mentioned, we set the detuning parameters equal to zero, and assume 
that the coupling of
the field to the atom is a constant independent of the intensity of the field, 
(i.e., $R = a$). 
For the sake of clarity, we outline below  the salient steps in obtaining the 
state 
of the 
system at a subsequent time, given the initial state. 

The eigenstates of $H_{0}$ are chosen as a basis, while $H_{1}$ is
treated as an interaction Hamiltonian.
Throughout this work,  we take  the initial state of the atom 
to be $\ket{1}$, and that of the field 
to be some specified superposition $\sum_{n=0}^{\infty} q_{n}\ket{n}$
of $n$-photon (Fock) states.
We denote by $\ket{j;n}$ the direct product state in which the atom is in state 
$\ket{j} \;(j = 1,2,3)$ and 
the field is in an $n$-photon state. 
Then (retaining the general notation of \cite{zait2003nonclassical}),
the  state of the system in the
interaction picture at any time $t \geq 0$ is of the form
\begin{eqnarray}
\fl \ket{\psi(t)}_{\rm int} = \sum\limits_{n=0}^{\infty} q_{n}\biggl\{ A_{n}(t)
 \ket{1;n} + B_{n}(t) \ket{2;n} + \nonumber \\
C_{n+1}(t) \ket{3;n+1} \biggr\}.
\label{eqn:single_mode_v_state_t}
\end{eqnarray}
The time-dependent coefficients 
$A_{n}(t)$, $B_{n}(t)$ and $C_{n+1}(t)$ satisfy the coupled differential equations 
\numparts
\begin{equation}
i \dot{A}_{n}  =  V_{1} A_{n}  +  f_{1}C_{n+1}
\label{eqn:single_mode_v_adot}
\end{equation}
\begin{equation}
i \dot{B}_{n}  =  V_{1}B_{n}  +  f_{2} C_{n+1}
\label{eqn:single_mode_v_bdot}
\end{equation}
\begin{equation}
i \dot{C}_{n+1}  =  V_{2} C_{n+1}  +  f_{1} A_{n}  +  f_{2} B_{n},
\label{eqn:single_mode_v_cdot}
\end{equation}
\endnumparts
where  
\numparts
\begin{equation}
V_{1} = \chi n (n-1), \quad V_{2} = \chi n (n+1)
\label{eqn:single_mode_v_v1_v2}
\end{equation}
\begin{equation}
f_{1} = \lambda_{1} \sqrt{n+1}, \quad f_{2} = \lambda_{2} \sqrt{n+1}.
\label{eqn:single_mode_v_f1_f2}
\end{equation}
\endnumparts
The trial solution $B_{n}\sim e^{i \mu t}$ gives us the following cubic equation 
for the characteristic frequency $\mu$:
\begin{equation}  
\mu^{3}  +  x_{1} \mu^{2}  +  x_{2} \mu  +  x_{3}  =  0,
\label{eqn:mu_cubic}
\end{equation}
where
\numparts
\begin{equation}
x_{1}  =  2 V_{1}  +  V_{2}
\label{eqn:single_mode_v_x1}
\end{equation}
\begin{equation}
x_{2}  =  V_{1} ( 2 V_{2}  +  V_{1} )  -  f_{1}^{2}  -  f_{2}^{2}
\label{eqn:single_mode_v_x2}
\end{equation}
\begin{equation}
x_{3}  = - V_{1} ( V_{2} V_{1} -  f_{1}^{2}  -  f_{2}^{2} ).
\label{eqn:single_mode_v_x3}
\end{equation}
\endnumparts
This equation has three
real roots given by
\begin{eqnarray}
\fl \mu_{j}  =  - \frac{x_{1}}{3}  +  \frac{2}{3} 
\sqrt{(x_{1}^{2}
-  3 x_{2})} \,\cos{\left\{\theta  +  
\frac{2}{3} (j-1) \pi  \right\}} , \nonumber \\
\space (j  =  1,2,3)
\label{eqn:mu_j}
\end{eqnarray}
where
\begin{equation}
\theta  =  \frac{1}{3}  \cos^{-1}{ \left\{ \frac{9 x_{1} x_{2}
-  2 x_{1}^{3}  -  27 x_{3}}{2 ( x_{1}^{2}  -  3 x_{2} )^{\frac{3}{2}} }
\right\}}.
\label{eqn:theta}
\end{equation}
Substituting $B_{n}(t)  =  \sum\limits_{j=1}^{3} b_{j} e^{i \mu_{j} t}$ in
\eref{eqn:single_mode_v_adot}--\eref{eqn:single_mode_v_cdot} gives
\numparts
\begin{equation}
A_{n}(t)  =  \sum\limits_{j=1}^{3} \frac{b_{j}}{f_{2}f_{1}} \left\{ ( \mu_{j}  +  
V_{1} 
)
( \mu_{j}  +  V_{2} )  -  f_{2}^{2} \right\} e^{i \mu_{j} t},
\label{eqn:single_mode_v_a}
\end{equation}
\begin{equation}
C_{n+1}(t)  =  - \sum\limits_{j=1}^{3} \frac{b_{j}}{f_{2}} ( \mu_{j}
+  V_{1} ) e^{i \mu_{j} t}.
\label{eqn:single_mode_v_c}
\end{equation}
\endnumparts
The coefficients $b_{j}$ are given by
\begin{equation}
b_{j}  =  \frac{f_{1} f_{2}}{(\mu_{j}- \mu_{k}) (\mu_{j}- \mu_{l})}, 
\label{eqn:b_j}
\end{equation}  
where, for each $j$, the indices $k$ and $l$ take the other two distinct values.
 
We now have the complete state of the system at time $t$. The density matrix
$\rho (t)$ for the system can now be contructed,  and the reduced 
density matrices
$\rho_{A} (t)$ and $\rho_{F} (t)$ for the atom and field respectively can be 
obtained. These are given by
\numparts 
\begin{equation}
\bra{1} \rho_{A} (t) \ket{1}  =  \sum\limits_{n=0}^{\infty} q_{n} 
q_{n}^{*} A_{n} A_{n}^{*}
\end{equation}
\begin{equation}
\bra{2} \rho_{A} (t) \ket{2}  =  \sum\limits_{n=0}^{\infty} q_{n} 
q_{n}^{*} B_{n} B_{n}^{*}
\end{equation}
\begin{equation}
\bra{3} \rho_{A} (t) \ket{3}  =  \sum\limits_{n=1}^{\infty} q_{n-1} 
q_{n-1}^{*} C_{n} C_{n}^{*}
\end{equation}
\begin{equation}
\bra{1} \rho_{A} (t) \ket{2}  =  \sum\limits_{n=0}^{\infty} q_{n} 
q_{n}^{*} A_{n} B_{n}^{*}
\end{equation}
\begin{equation}
\bra{1} \rho_{A} (t) \ket{3}  =  \sum\limits_{n=1}^{\infty} q_{n} 
q_{n-1}^{*} A_{n} C_{n}^{*}
\end{equation}
\begin{equation}
\bra{2} \rho_{A} (t) \ket{3}  =  \sum\limits_{n=1}^{\infty} q_{n} 
q_{n-1}^{*} B_{n} C_{n}^{*},
\end{equation}
\endnumparts
and
\begin{eqnarray}
\fl \bra{n} \rho_{F} (t) \ket{n^{\prime}}  =  q_{n} 
q_{n^{\prime}}^{*} ( A_{n} A_{n^{\prime}}^{*}  +  
B_{n} B_{n^{\prime}}^{*} ) \nonumber \\
+(1 - \delta_{n,0})(1 - \delta_{n', 0}) 
q_{n-1} q_{n^{\prime}-1}^{*} C_{n} C_{n^{\prime}}^{*}.
\label{eqn:single_mode_v_rho_f_matrix_elts}
\end{eqnarray}
Using these results, we now proceed to investigate numerically the temporal 
behaviour of the SVNE when the 
initial 
state of the 
radiation field is either a CS or a PACS. 
We have also computed the
Mandel $Q$ parameter as a function of time, for these initial states.
For simplicity, we have set $\lambda_{1} = \lambda_{2} = \lambda$ 
(so that $f_{1} = f_{2}$) in the numerics.
As already stated, we work in a parameter regime  
where the nonlinearity in the field dominates over the atom-field interaction, 
by taking the ratio $\chi/\lambda$ to be $5$. 
Consistent with the notation $\ket{j;n}$, we denote 
by $\ket{1; \alpha}$ a state in which the atom is in state $\ket{1}$ and the field 
is in a CS $\ket{\alpha}$. Similarly,  
$\ket{1; \alpha,m}\; (m = 1,2, \ldots)$ denotes a state in which the field is 
in an $m$-photon added PACS. 

In figures 1(a)--(c) we have plotted the SVNE corresponding to the field 
subsystem, $- {\rm Tr}\; (\rho_{F}\, \log_{2}\, \rho_{F})$, 
as a function of the scaled time
$\lambda t$, for different initial states of 
the field.  Note  the different time scales in figure 
1(a) and in figures 1(b), (c). 
It is seen from figure 1(a) that an initially unentangled state $\ket{1; \alpha}$ 
(with SVNE equal to zero) 
gets entangled over a very short time. Although this initial state does not revive 
fully at 
any later time, the SVNE drops to relatively small values almost periodically 
(when $\lambda t \approx 400, 800, \ldots $). 
In between these times 
marked oscillatory behaviour is seen in the SVNE with less-pronounced dips close to 
instants 
of 
approximate fractional revivals. Thus, signatures of  near-revival phenomena  
are captured in the SVNE in this system, even for the case of strong nonlinearity. 
This 
is to be contrasted with  earlier results  
\cite{sudhjphysb} on the system in \cite{puri} where, 
regardless of the  initial state,  
plausible signatures of 
near-revival phenomena are seen in the SVNE only in the weak 
nonlinearity regime. 
The finite-dimensional Hilbert space of the V-type atom in the 
present model 
plays a crucial role in this regard. 
As in the cases reported in \cite{sudhjphysb}, any significant departure of the 
initial field state from coherence, or an initial CS with large  
${|\alpha|}^{2}$,
essentially erases even near-revivals. This is seen in figures 1(b) and (c): 
the SVNE never gets 
close to its initial value of zero at a later time.  

Next, we turn to the Mandel $Q$ parameter, defined as 
\begin{equation}
Q = \frac{\langle (a^{\dagger}a)^{2} \rangle - {\langle a^{\dagger} a
\rangle}^{2}}{\langle a^{\dagger} a \rangle} - 1
\label{eqn:mandel_q}
\end{equation}
$Q$ is a measure of the deviation of the photon number distribution from Poisson 
statistics (which is, of course, a hallmark of a CS).
Figures 2(a)--(c) show  $Q$ as a 
function of time, for the same 
initial states and parameter values as above. Once again, the time scale in each 
case is 
selected appropriately for the sake of clarity. 
It is clear from figure 2(a) that for an initial CS with 
sufficiently small ${|\alpha|}^{2}$ ( taken to be $1$ in this case), $Q$ 
is roughly oscillatory in time, and takes  values ranging from 
$\approx -0.35$ to $\approx +0.05$, during the time interval considered. 
This behaviour is 
suggestive of 
sub-Poissonian statistics over a significant period of time. However, the field also 
evolves periodically through states displaying Poisson and marginally super-Poisson 
statistics. Taken together with the behaviour of the SVNE in figure 1(a), it is 
clear that 
this oscillatory behaviour reflects the quasiperiodic nature of the dynamics in this 
case. 

In contrast to this, it is evident from figure 2(b) that for the 
same value 
of  ${|\alpha|}^{2}$,  an  
initial field state 10-PACS (which exhibits sub-Poissonian number statistics),  
evolves through several states which exhibit the same photon number statistics for a 
significant 
part of its temporal evolution, with the value of $Q$ always 
remaining close to its initial value. Interestingly, however, thee are essentially 
no near-revivals in this case, as is clear from figure 1(b).
A similar 
inference may be drawn from figures 1(c) and 2(c), representing the case in 
which the initial 
state of 
the field is a CS with ${|\alpha|}^{2} = 10$. Its subsequent evolution appears to be 
through states with marginally sub-Poissonian statistics for a significant part of 
the time. 

\begin{figure}[H] 
\centering \includegraphics[scale=0.5]{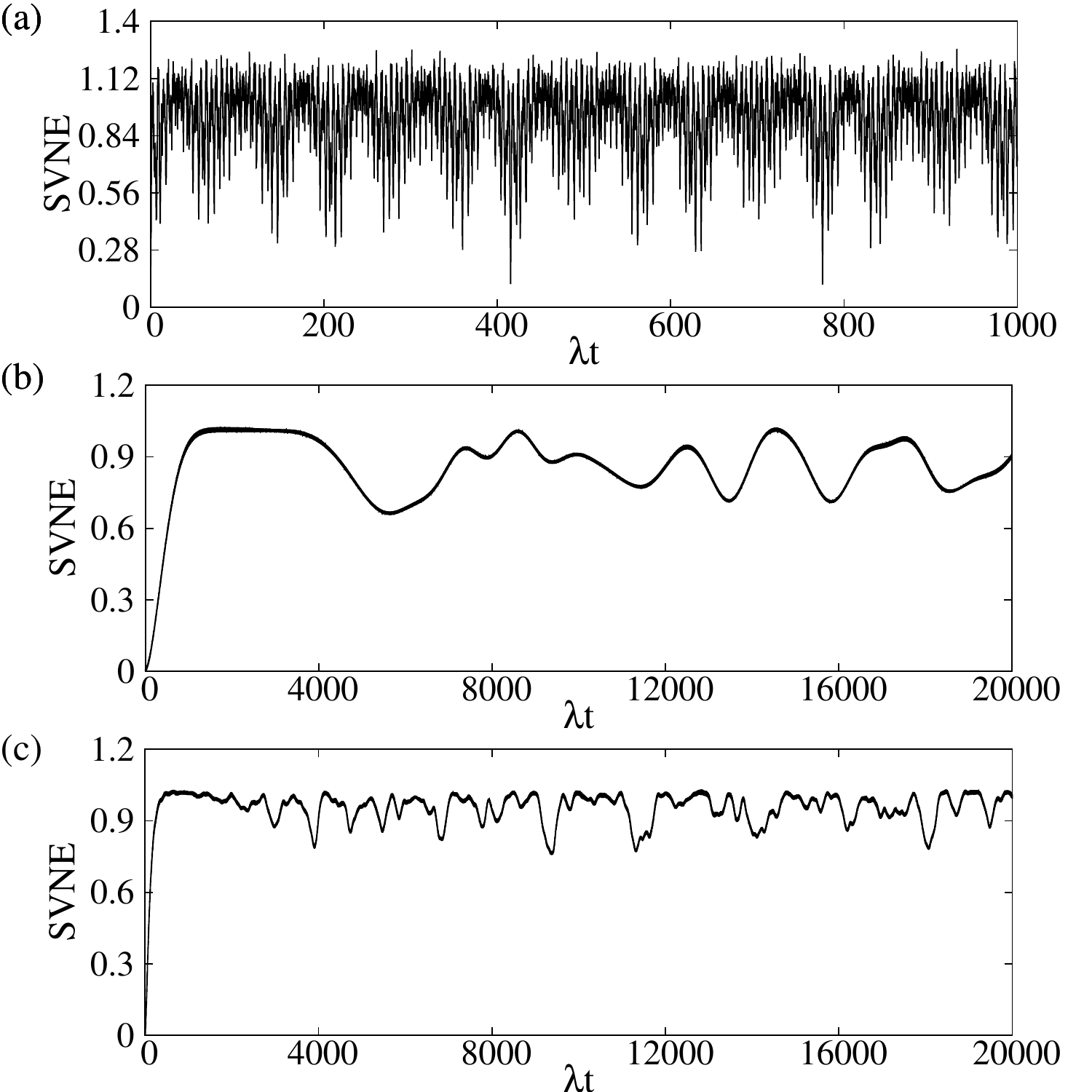}
\caption{SVNE as a function of the scaled time $\lambda t$ for the three-level 
V-type 
atom interacting with a single-mode field, for strong nonlinearity 
($\chi/\lambda = 5$). The initial states are (a) $\ket{1;\alpha}$, ${|\alpha|}^{2} = 
1$, (b) $\ket{1;\alpha,10}$, ${|\alpha|}^{2} =1$, and (c) $\ket{1;\alpha}$, 
${|\alpha|}^{2} = 10$.}
\label{fig:single_mode_v_svne}
\end{figure}

\begin{figure}[H]
\centering
\includegraphics[width=0.5\textwidth]{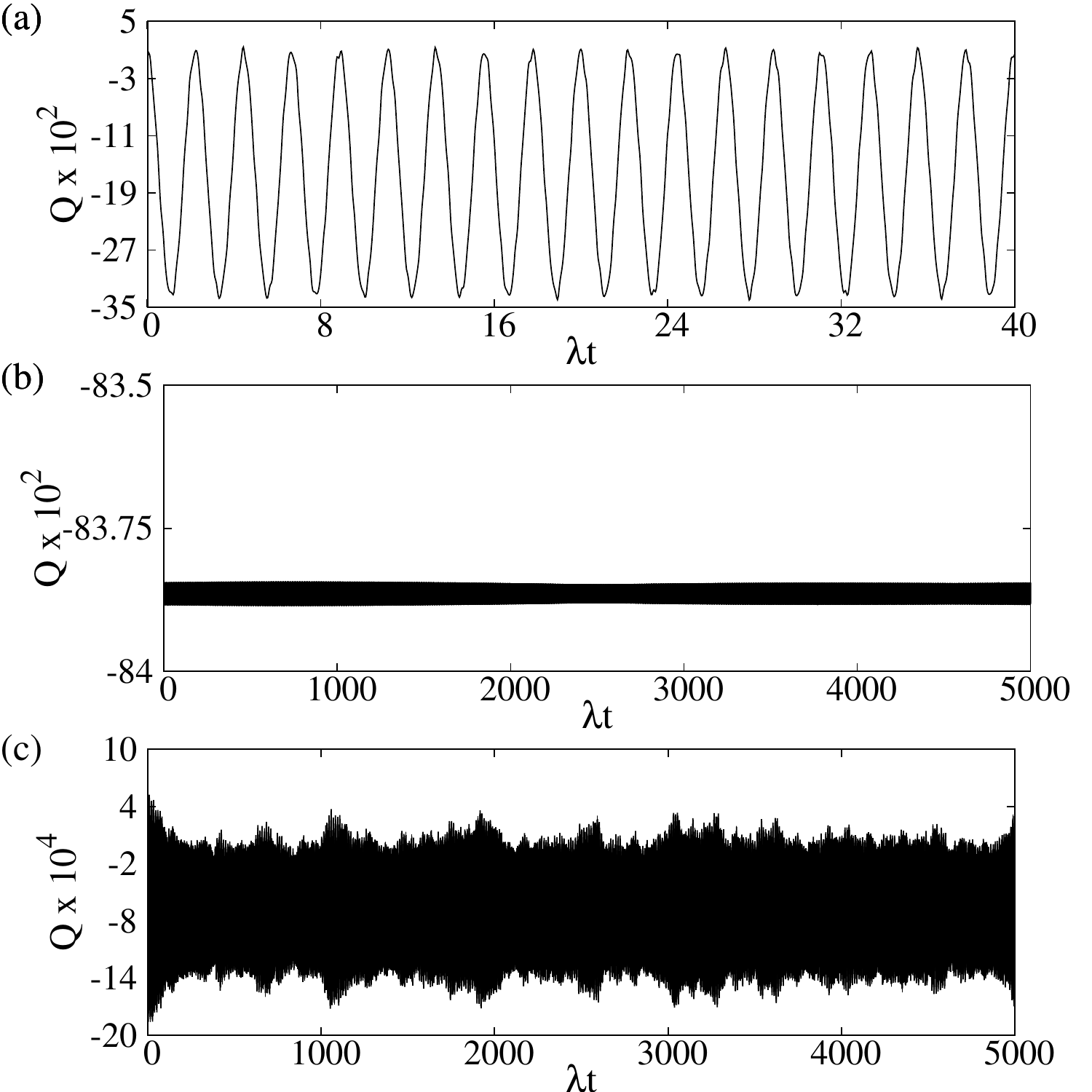}
\caption{Mandel $Q$ parameter as a function of the scaled time $\lambda t$ for the 
three-level V-type 
atom interacting with a single-mode field,  for strong nonlinearity
($\chi/\lambda = 5$). The initial states are (a) $\ket{1;\alpha}$, ${|\alpha|}^{2} =
1$, (b) $\ket{1;\alpha,10}$, ${|\alpha|}^{2} =1$, and (c) $\ket{1;\alpha}$,
${|\alpha|}^{2} = 10$.}
\label{fig:single_mode_v_mandel_q}
\end{figure}


\subsection{Interaction with two field modes}

We now consider a three-level V-type atom interacting with two field modes, 
of respective frequencies $\Omega_{1}$ and $\Omega_{2}$ and associated operators 
$(a_{1}, a_{1}^{\dagger})$ and $(a_{2}, a_{2}^{\dagger})$. We designate these modes 
as $F_{1}$ and $F_{2}$, for convenience. They induce, respectively, 
 $\ket{3} \leftrightarrow \ket{1}$ and  $\ket{3} \leftrightarrow \ket{2}$
transitions. 
The general Hamiltonian which incorporates nonlinearity in the field 
modes, intensity-dependent couplings 
and non-zero 
detuning parameters is an extension of  
(\ref{eqn:single_mode_v_hamiltonian}), 
and is of the form
\begin{eqnarray}
\fl \tilde{H}  =  \sum\limits_{j=1}^{3} \omega_{j} \sigma_{jj}  +  
\Omega_{1} a_{1}^{\dagger} a_{1}  +  \chi_{1} a_{1}^{\dagger 2} 
a_{1}^{2}  +  
\Omega_{2} a_{2}^{\dagger} a_{2}  +  \chi_{2} a_{2}
^{\dagger 2} a_{2}^{2} \nonumber \\
 +  \lambda_{1} ( R_{1} \sigma_{13}  
+  R^{\dagger}_{1} \sigma_{31} )  +  
\lambda_{2} ( R_{2} \sigma_{23}  
+  R^{\dagger}_{2} \sigma_{32} )
\label{eqn:two_mode_v_hamiltonian}
\end{eqnarray}
where  $R_{1}  =  a_{1} f_{1}(N_{1})$ and $R_{2}  
=  a_{2} f_{2}(N_{2})$. 
As in the bipartite case, we will examine the
temporal evolution of
the SVNE and the $Q$ parameter corresponding to various initial states of
subsystems $F_{1}$ and $F_{2}$. 

Once again, we write $ \tilde{H} =\tilde{H}_{0} +  \tilde{H}_{1}$,
where $[\tilde{H}_{0} , \tilde{H}_{1}] = 0$ and 
\numparts
\begin{eqnarray}
\tilde{H}_{0}  =  \omega_{3} I  +  \Omega_{1} N_{1}^{tot}  +  \Omega_{2} 
N_{2}^{tot},
\label{eqn:two_mode_v_h0}
\end{eqnarray}
\begin{eqnarray}
\fl \tilde{H}_{1}  =   \chi_{1} a_{1}^{\dagger 2} a_{1}^{2}  +  
\chi_{2} a_{2}^{\dagger 2} a_{2}^{2}  -  \Delta_{1} \sigma_{11} 
 -  \Delta_{2} \sigma_{22} \nonumber \\  
+  \lambda_{1} ( R_{1} \sigma_{13}  +  R^{\dagger}_{1} 
\sigma_{31} ) 
 +  \lambda_{2} ( R_{2} \sigma_{23}  
+  R^{\dagger}_{2} \sigma_{32} ).
\label{eqn:two_mode_v_h1}
\end{eqnarray}
\endnumparts
As before, $I  =  \sum\limits_{j=1}^{3} \sigma_{jj}$, 
$N_{1}^{tot}  =  a_{1}^{\dagger} a_{1}  +  \sigma_{11}$,
$N_{2}^{tot}  =  a_{2}^{\dagger} a_{2}  +  \sigma_{22}$, and
the detuning parameters are 
\numparts
\begin{equation}
\Delta_{1} = \omega_{3} - \omega_{1} + \Omega_{1}
\label{eqn:two_mode_v_delta1}
\end{equation}
\begin{equation}
\Delta_{2} = \omega_{3} - \omega_{2} + \Omega_{2}.
\label{eqn:two_mode_v_delta2}
\end{equation}
\endnumparts
We have computed the time-dependent density matrix for a generic initial state 
governed by this 
Hamiltonian. 
For the purpose at hand, however, we restrict ourselves to results in the 
case of zero detuning and intensity-independent couplings. $F_{1}$ and $F_{2}$ are 
in initial states given respectively by $\sum\limits_{n=0}^{\infty} q_{n} \ket{n}$ 
and $\sum\limits_{m=0}^{\infty} r_{m}
\ket{m}$, and the atom is taken to be in the state $\ket{1}$.
In an obvious extension of the notation already used, 
\begin{equation}
\ket{\psi (0)}  =  \sum\limits_{n=0}^{\infty} \sum\limits_{m=0}^{\infty} q_{n} r_{m} 
\ket{1; n; m}. 
\label{eqn:two_mode_v_initial_state}
\end{equation}  
The state of the system 
at time $t$  
in the interaction picture is of the form
\begin{eqnarray}
\fl \ket{\psi (t)}_{int} = \sum\limits_{n=0}^{\infty}\sum\limits_{m=0}^{\infty} 
q_{n} r_{m} \biggl\{ A_{nm}(t) \ket{1; n; m}
 \nonumber \\
 + B_{nm}(t) \ket{2; n+1; m-1} \nonumber \\
 + C_{nm}(t) \ket{3; n+1; m} \biggr\}.
\label{eqn:two_mode_v_interaction_state}
\end{eqnarray}
It is to be understood implicitly that $B_{n0} = 0$.

First, we consider the case $m \geq 1$. It is helpful to define
\numparts
\begin{equation}
V_{11}  =  \chi_{1} n (n-1), \quad V_{12}  =  \chi_{1} n (n+1)
\label{eqn:two_mode_v_v11v12}
\end{equation}
\begin{equation}
V_{21}  =  \chi_{2} (m-1) (m-2), \quad V_{22}  =  \chi_{2} m (m-1)
\label{eqn:two_mode_v_v21v22}
\end{equation}
\begin{equation}
f_{1}  =  \lambda_{1} \sqrt{n+1}, \quad f_{2}  =  \lambda_{2} \sqrt{m}.
\label{eqn:two_mode_v_f1f2}
\end{equation}
\endnumparts
Then, the coupled differential equations 
for the coefficients  $A_{nm}(t)$, $B_{nm}(t)$ and $C_{nm}(t)$ are 
given by 
\numparts
\begin{equation}
i \dot{A}_{nm}  =  (V_{11} + V_{22})A_{nm}  +   f_{1} C_{nm}
\label{eqn:two_mode_v_adot}
\end{equation}
\begin{equation}
i \dot{B}_{nm}  =  (V_{12} + V_{21})B_{nm}  +   f_{2} C_{nm}
\label{eqn:two_mode_v_bdot}
\end{equation}
\begin{equation}
i \dot{C}_{nm}  =  (V_{12} + V_{22})C_{nm} +   f_{1} A_{nm}  +  f_{2} B_{nm}.
\label{eqn:two_mode_v_cdot}
\end{equation}
\endnumparts
Once again, the trial solution $B_{nm} \sim e^{i \mu t}$ yields the cubic equation 
(\ref{eqn:mu_cubic}) for the characteristic frequency $\mu$, 
but with the coefficients
\numparts
\begin{eqnarray}
x_{1}  =  V_{11}  +  2 V_{12}  +  V_{21}  +  2 V_{22},
\label{eqn:two_mode_v_x1}
\end{eqnarray}
\begin{eqnarray}
\fl x_{2}  =  (V_{12}  +  V_{21}) (V_{11}  +  V_{12}  +  2 V_{22}) \nonumber \\ 
+  (V_{12}  +  V_{22}) (V_{11}  +  V_{22})  - f_{1}^{2}  -  f_{2}^{2},
\label{eqn:two_mode_v_x2}
\end{eqnarray}
\begin{eqnarray}
\fl x_{3}  =  (V_{12}  +  V_{21}) \left\{ (V_{12}  +  V_{22}) (V_{11}  
+  V_{22})  -  f_{1}^{2} \right\} \nonumber \\
-  f_{2}^{2} (V_{11}  + V_{22}).
\label{eqn:two_mode_v_x3}
\end{eqnarray}
\endnumparts
As before, the solutions $\mu_{j} (j = 1,2,3)$ of the cubic equation  
are of the form written down 
in  \eref{eqn:mu_j}.
The solutions for the coefficients are 
\numparts
\begin{equation}
A_{nm}(t)  =  \frac{1} {f_{1}f_{2}} \sum\limits_{j=1}^{3} b_{j} \left\{ (\mu_{j}  +  
V_{12}  
+  V_{22}) (\mu_{j}  +  V_{12}  +  V_{21})  -  f_{2}^{2} \right\}  e^{i \mu_{j} t}
\label{eqn:two_mode_v_a}
\end{equation}
\begin{equation}
B_{nm}(t)  =  \sum\limits_{j=1}^{3} b_{j} e^{i \mu_{j} t}
\label{eqn:two_mode_v_b}
\end{equation}
\begin{equation}
C_{nm}(t)  = -  \frac{1} {f_{2}} \sum\limits_{j=1}^{3} b_{j} (\mu_{j}  +  V_{12}  
+  V_{21}) e^{i \mu_{j} t},
\label{eqn:two_mode_v_c}
\end{equation}
\endnumparts
where $b_{j}$ has the same form as in 
(\ref{eqn:b_j}).

Next, consider the case $m = 0$, in 
which the only 
possibilities are
$\ket{3} \leftrightarrow \ket{1}$ transitions. 
Equations \eref{eqn:two_mode_v_adot} and \eref{eqn:two_mode_v_cdot} now reduce 
to
\numparts
\begin{equation}
i \dot{A}_{nm}  =  V_{11} A_{nm}  +  f_{1} C_{nm}
\label{eqn:two_mode_v_adot_m0}
\end{equation}
\begin{equation}
i \dot{C}_{nm}  =  V_{12} C_{nm}  +  f_{1} A_{nm}.
\label{eqn:two_mode_v_cdot_m0}
\end{equation}
\endnumparts
Setting $A \sim e^{i \alpha t}$ yields the secular equation
\begin{equation}
\alpha^{2}  +  y_{1} \alpha  +  y_{2}  =  0
\label{eqn:two_mode_v_alpha_quad}
\end{equation}
for the characteristic frequency $\alpha$,
where
\numparts
\begin{equation}
y_{1}  =  V_{11}  +  V_{12}
\label{eqn:two_mode_v_y1}
\end{equation}
\begin{equation}
y_{2}  =  V_{12} V_{11}  -  f_{1}^{2}.
\label{eqn:two_mode_v_y2}
\end{equation}
\endnumparts
Denoting the solutions of (\ref{eqn:two_mode_v_alpha_quad})  by 
$\alpha_{1}$ and $\alpha_{2}$, 
and using the fact that the initial state of the atom is $\ket{1}$ 
(so that $A_{nm}(0) = 1, B_{nm}(0) = C_{nm}(0) = 0)$, 
we obtain 
\numparts
\begin{equation}
A_{nm}(t) = \sum\limits_{j=1}^{2} c_{j} e^{i \alpha_{j} t}
\label{eqn:two_mode_v_a_m0}
\end{equation}
\begin{equation}
C_{nm}(t) = - \frac{1} {f_{1}} \sum\limits_{j=1}^{2} c_{j} (\alpha_{j}  +  V_{11}) 
e^{i \alpha_{j} t},
\label{eqn:two_mode_v_c_m0}
\end{equation}
\endnumparts
where 
\numparts
\begin{equation}
c_{1}  =  \frac{V_{11} + \alpha_{2}} {\alpha_{2}  -  \alpha_{1}}
\label{eqn:two_mode_v_a1}
\end{equation}
\begin{equation}
c_{2}  =  \frac{V_{11} + \alpha_{1}} {\alpha_{1}  -  \alpha_{2}}.
\label{eqn:two_mode_v_a2}
\end{equation}
\endnumparts
We now have the complete solution for  $\ket{\psi (t)}_{int}$ in explicit form, 
from which we can 
construct 
the density matrix
$\rho (t)$ for the system and obtain the reduced density matrix $\rho_{F_{1}} (t)$
for the field subsystem $F_{1}$.  
The general matrix element of this quantity is given by 
\begin{eqnarray}
\fl \bra{n} \rho_{F_{1}} (t) \ket{n^{\prime}} & =& \sum\limits_{l=0}^{\infty} 
\bigg[ q_{n} 
q_{n^{\prime}}^{*} r_{l} r_{l}^{*} A_{n,l} A_{n^{\prime},l}^{*} \nonumber \\
&+& (1-\delta_{n,0})(1-\delta_{n^{\prime},0}) \nonumber \\ 
& \times & \big\{ q_{n-1} 
q_{n^{\prime}-1}^{*} 
r_{l+1} r_{l+1}^{*} B_{n-1,l+1} B_{n^{\prime}-1,l+1}
^{*} \nonumber \\
&+&  q_{n-1} q_{n^{\prime}-1}^{*} r_{l} r_{l}^{*} C_{n-1,l} C_{n^{\prime}-1,l}^{*} 
\big\} \bigg]
\label{eqn:two_mode_v_rho_f1_matrix_elts}
\end{eqnarray}
where the $t$-dependence of $A_{nm}, B_{nm}, C_{nm}$ has been suppressed for 
simplicity. 

Using the reduced density matrix found above, we have 
investigated in detail the temporal behaviour of the 
SVNE and the $Q$ parameter.   
The inferences we draw concerning  the dynamics turn out to be 
consistent 
with those obtained by treating the atom and the field mode $F_{2}$ 
together as the 
second subsystem of a bipartite system, and examining its dynamics. 
As before, we consider (for simplicity) the case 
$\lambda_{1} = \lambda_{2} = \lambda$ and  $\chi_{1} = \chi_{2} = \chi$.  
We set $\chi / \lambda =5$, in order to 
facilitate
comparison between the present case and the bipartite case considered 
earlier, as well as the case reported in \cite{sudhjphysb}.

Figures 3(a) and (b) depict the SVNE for 
the subsystem $F_{1}$ as a function of the scaled time $\lambda t$, for initially 
unentangled states $\ket{1;\alpha; \alpha}$ with ${|\alpha|}^{2} = 1$ and 
${|\alpha|}^{2} = 10$ respectively. It is evident (see figure 3(a)) that the 
tripartite case is bereft 
of even near-revivals, in the regime of strong nonlinearity, even for small 
values of ${|\alpha|}^{2}$.  This is in contrast to figure 1(a) (which also 
corresponds to 
${|\alpha|}^{2} = 
1$), and is akin to the 
situation reported in \cite{sudhjphysb}, where each of the two subsystems 
of the 
bipartite system had `large' Hilbert spaces, in contrast to the bipartite system 
considered earlier in this paper. This indicates that even if two subsystems of a
full system are 
significantly `large' (in the sense of a high-dimensional Hilbert space), 
the revival phenomenon is absent for sufficiently strong nonlinearities,  
independent of whether these large subsystems interact with each other 
directly or through  much smaller subsystems. 

For the same tripartite initial states as in figures 3(a) and (b), we have plotted 
the $Q$ parameter as a function of $\lambda t$ in figures 4(a) and (b). Once again 
we note that an initial coherent state of the subsystem evolves through a series of 
mixed states with sub-Poissonian number statistics for a significant part of the 
time, and through other mixed states also having  Poissonian  number statistics. The 
somewhat oscillatory nature of  $Q$  for small ${|\alpha|}^{2}$ as  
in figure 4(a) is replaced by bursts of oscillations in figure 4(b). The absence of 
revivals of the initial state $\ket{1;\alpha; \alpha}$ with  ${|\alpha|}^{2} = 10$, 
taken together with the absence of regular oscillatory behaviour of $Q$ 
for this state as it evolves in time, suggests that a detailed 
time series 
analysis of an appropriate observable such as the mean photon number would reveal a 
rich diversity of ergodicity properties. 
For ready comparison, a similar analysis is necessary 
for appropriate initial states of the bipartite model of the previous subsection.
These results are presented in the next section.

\begin{figure}[H]
\centering
\includegraphics[scale=0.5]{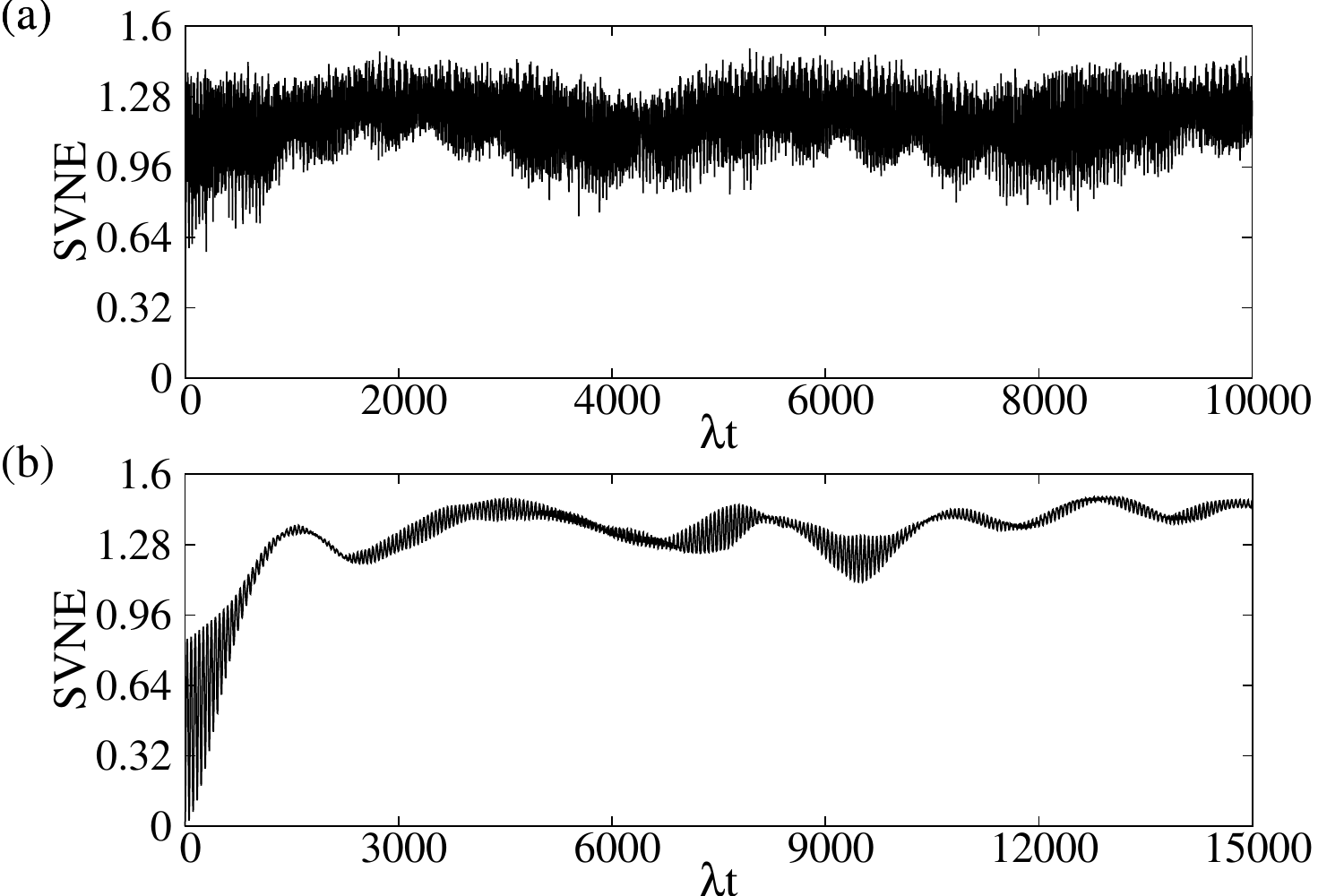}
\caption{SVNE as a function of scaled time $\lambda t$ for the three-level V-type
atom interacting with two fields modes, for strong nonlinearity
($\chi/\lambda = 5$) and initial states (a) $\ket{1;\alpha; \alpha}$, 
${|\alpha|}^{2} 
= 1$, and  (b) $\ket{1;\alpha; \alpha}$, ${|\alpha|}^{2} =10$.} 
\label{fig:two_mode_v_svne}
\end{figure}
\begin{figure}[H]
\centering
\includegraphics[width=0.5\textwidth]{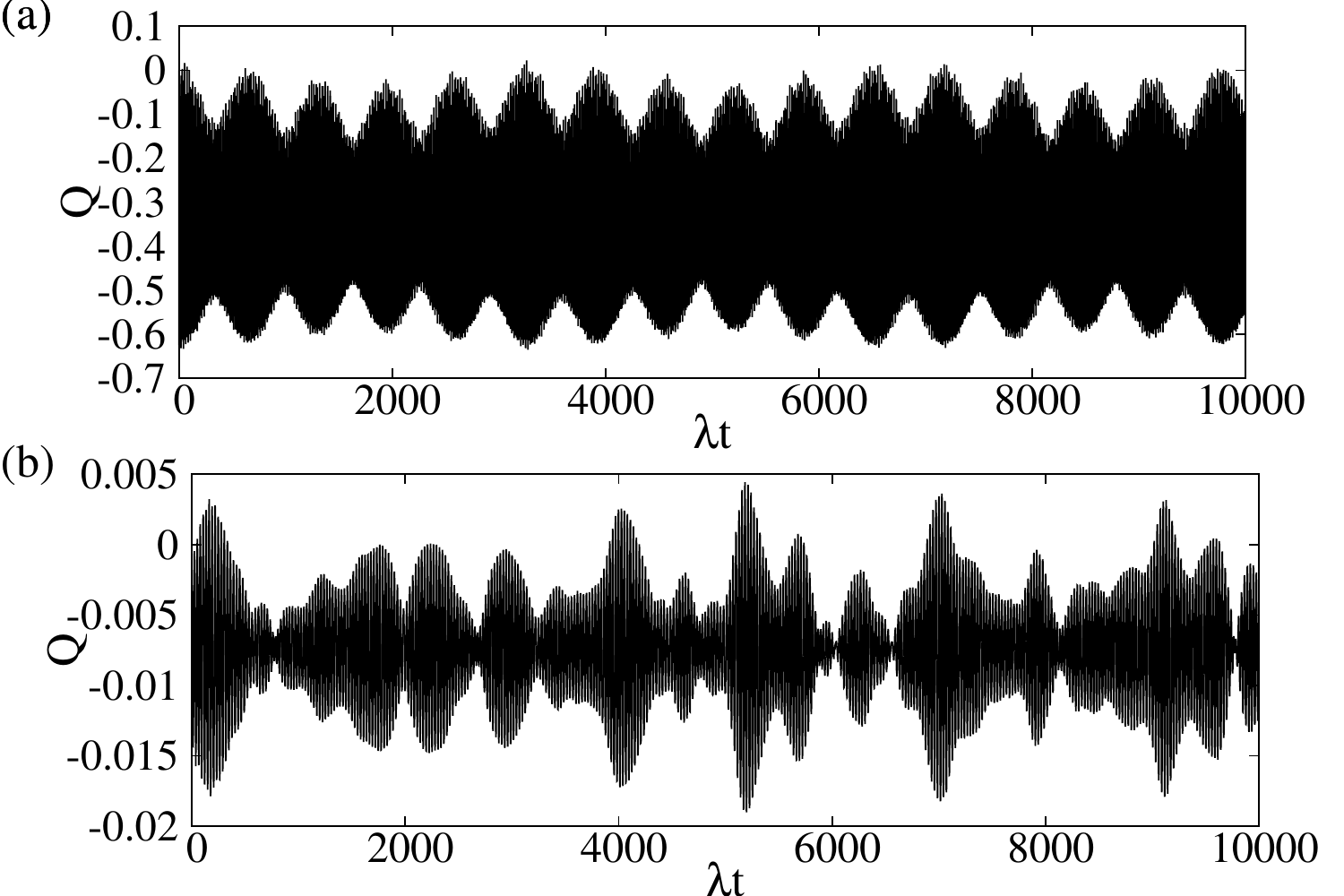}
\caption{Mandel $Q$ parameter as a function of scaled time $\lambda t$ for the 
three-level 
V-type
atom interacting with two fields modes, for strong nonlinearity
($\chi/\lambda = 5$) and initial states (a) $\ket{1;\alpha; \alpha}$,  
${|\alpha|}^{2} =
1$, and (b) $\ket{1;\alpha; \alpha}$, ${|\alpha|}^{2} =10$. 
}
\label{fig:two_mode_v_mandel_q}
\end{figure}


\section{Ergodicity properties of the mean photon number}
We are now in a position to investigate the dynamics of any appropriate observable 
in 
both the models described in Section 2.
In order to be specific, we  examine in detail the ergodicity 
properties of the mean photon 
number $\langle a^{\dagger}a \rangle$ 
of the radiation field as it interacts with the three-level atom in the bipartite 
model, and  the mean photon number $\langle a_{1}^{\dagger}a_{1} \rangle$
of $F_{1}$ in the tripartite model. The procedure we adopt is as 
follows. Starting with a specified initial state of the full system,
we generate a 
sufficiently long time 
series of the mean photon number (approximately $10^{7}$ data points obtained in 
time 
steps $\delta t$). The time in units of $\delta t$ is  denoted 
by $\tau$. The interval of values of the mean photon number is 
coarse-grained into small equal-sized cells, and the distribution of the time 
of the first 
recurrence to a given generic cell is determined. The plot of the 
observable at time $ \tau + 1$ versus its value at time  
${\tau}$ (the return map) has been obtained for different initial states. 
Further, we have carried 
out a detailed time series analysis including estimation 
of the minimum embedding dimension, phase space reconstruction, 
 and determination 
of the maximum Lyapunov exponent.   

Before presenting our results, we summarize briefly the inferences drawn in 
\cite{sudhrecurrence} based on a similar study undertaken on the bipartite model of 
\cite{puri}. There, too, the observable whose time series was analyzed was the mean 
photon 
number. The first return distributions for various initial coherent states  with 
small ${|\alpha|}^{2}$, were spiky, 
and the corresponding Lyapunov exponents were  zero, signalling regular 
quasiperiodic 
behaviour \cite{nicolis1} of the observable concerned. 
In contrast, in all cases where the initial state was a CS with large 
${|\alpha|}^{2}$ or a PACS, the
first return dustribution was exponential and the 
Lyapunov exponent was positive,  thereby indicating exponentially 
unstable dynamics. 
Both subsystems in this model had infinite-dimensional 
Hilbert spaces, in contrast 
to the two models  considered in the
present work, where the atomic transitions are confined 
to just  three levels and the 
two radiation fields in the tripartite model do not 
directly interact with each other. There is therefore 
no reason to expect, \`a priori,  
that the inferences drawn in \cite{sudhrecurrence} will hold good here.  

We start with the bipartite system of the 
three-level atom interacting with a single 
field mode. The first task is to  verify if the 
dynamics is metrically transitive 
for generic cells in 
the coarse-grained phase space of the mean photon number. 
Figure 5 is a plot of 
the mean recurrence time versus cell size for 
an initial state $\ket{1; \alpha}$ 
with ${|\alpha|}^{2} = 10$. We have used 
$10^{7}$ data points with a time step 0.005. 
The excellent linear fit confirms that the Poincar\'e
recurrence theorem is satisfied, confirming that the dynamics is ergodic.

\begin{figure}[H]
\centering
\includegraphics[width=0.5\textwidth]{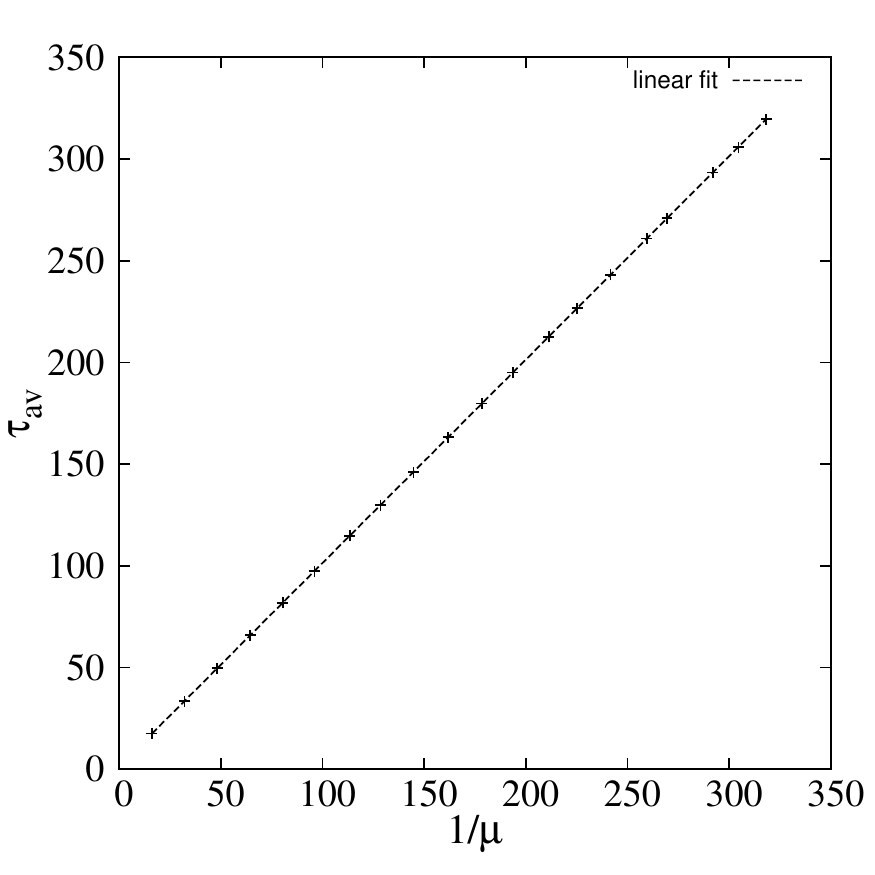}
\caption{Mean recurrence time of the mean photon number 
to a cell versus the reciprocal 
of its invariant measure, for the bipartite system.  
Initial state  $\ket{1; 
\alpha}$, with 
${|\alpha|}^{2} = 1$.}
\label{fig:single_mode_v_case13_poincare}
\end{figure}
 
Figure 6(a) shows the spiky first-return-time 
distribution $\phi_{1}$ to a generic cell for an 
initial state $\ket{1; \alpha}$ for ${|\alpha|}^{2} = 1$. Its spiky nature is an 
indication that the dynamics is quasiperiodic with several incommensurate
frequencies \cite{nicolis1}. 
With an  increase in ${|\alpha|}^{2}$ to the value 10, the 
distribution becomes exponential, as seen in  figure 6(b). 
The latter is the distribution 
expected for a hyperbolic dynamical system for a 
sufficiently small cell size 
\cite{hirata, balakrishnan}. We have further confirmed in this 
case, from the 
distributions for two, three 
and four  
successive recurrences to a generic cell, that 
such returns are uncorrelated, 
being given by the successive terms of a 
Poisson distribution \cite{hirata1, 
balakrishnan1}. 

We have also carried out a 
detailed time-series analysis of the mean photon number 
for the initial  states $\ket{1;\alpha}$ with $|\alpha|^{2} = 1$ and 
 $\ket{1;\alpha, 1}$ with $|\alpha|^{2} = 10$,   
with respective time steps of 0.25 
and 0.005. 
This study comprises phase space reconstruction, 
estimation of the minimum embedding dimension by 
the false-nearest-neighbours (FNN)  
algorithm \cite{kennel,abar}, and calculation of the maximal 
Lyapunov exponent using a robust algorithm by 
Rosenstein et al. \cite{rosenstein}. The algorithms 
have been implemented with the 
package TISEAN \cite{tisean}. In both 
cases ($|\alpha|^2 = 1$ and $|\alpha|^2 = 10, m = 1$) the 
Lyapunov 
exponent is found to be zero. While this is consistent
with the conclusion arrived at in  
\cite{sudhrecurrence} for a spiky first return distribution (Figure 6(a)), 
it is rather surprising in the case of an exponential first 
return distribution (which obtains for the initial state $\ket{1;\alpha, 1}$ with 
$|\alpha|^{2} = 10$),  as the latter is 
customarily associated with 
hyperbolicity. Evidently, the 
limited nature of the transitions allowed for the atomic subsystem in this 
case plays a crucial role in determining the ergodicity properties of the 
observable. 

\begin{figure}[H]
\centering
\includegraphics[width=\textwidth]{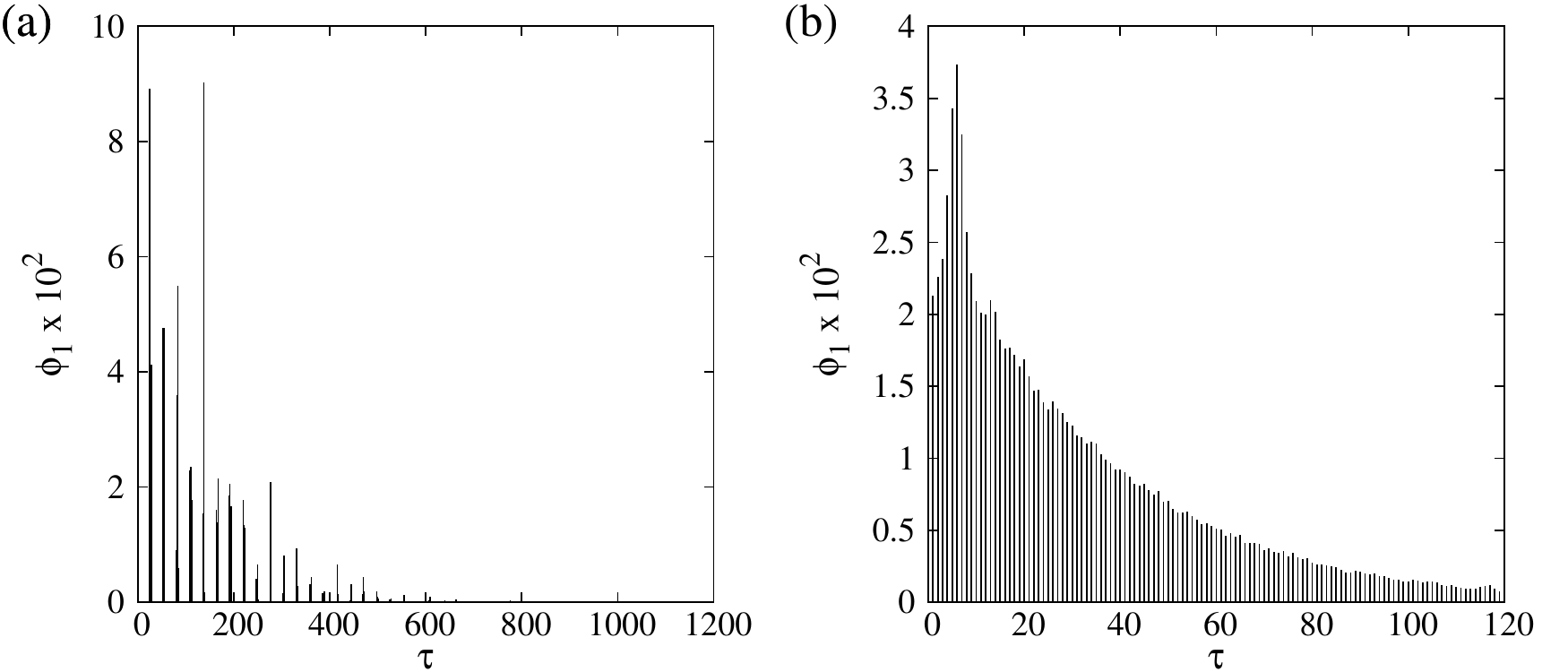}
\caption{First-return-time distribution $\phi_{1}$ of the mean photon number 
$\langle a^{\dagger}a \rangle$ for the 
bipartite system. Initial state  (a) $\ket{1; \alpha}$, ${|\alpha|}^{2} = 1$, and  
(b) $\ket{1; \alpha}$, ${|\alpha|}^{2} = 10$}
 \label{fig:single_mode_v_rtd_case7_case13}
\end{figure}

Figures 7(a)-(c) depict the return maps of the mean photon number in the 
bipartite model for 
initial states  $\ket{1; \alpha}$
with ${|\alpha|}^{2} = 1$, $\ket{1; \alpha, 10}$, 
${|\alpha|}^{2} = 1$, and  $\ket{1; \alpha}$
with ${|\alpha|}^{2} = 10$ respectively.  We have used 
$3 \times 10^{5}$ data points for 
generating these maps, with a time step 0.25 in figure 7(a) and 0.005 
in figures 
7(b) and (c). As 
expected, for an initial CS with a small 
value of ${|\alpha|}^{2}$ (figure 7(a)), the return map is an annulus 
with no 
sub-structure. For an initial PACS with the same small 
value of $|\alpha|^{2}$, the 
annulus is considerably 
broadened, and has sub-structures in it (figure 7(b)).
These features are consistent with corresponding first-return 
distributions that are spiky and vanishing Lyapunov exponents.
In contrast, the return map (figure 7(c)) for an 
initial CS with large ${|\alpha|}^{2}$ is seen to tend towards
the space-filling map 
reminiscent of a ` chaotic'   
system, although the Lyapunov exponent remains zero in this case.
We therefore conclude that, in  the presence of a subsystem  
with a low-dimensional
Hilbert space, exponential first return time distributions 
and space-filling return maps 
merely signal hyperbolicity without necessarily implying 
chaotic behaviour (as characterized by a  positive Lyapunov exponent).       
 
\begin{figure}[H]
\centering
\includegraphics[width=0.5\textwidth]{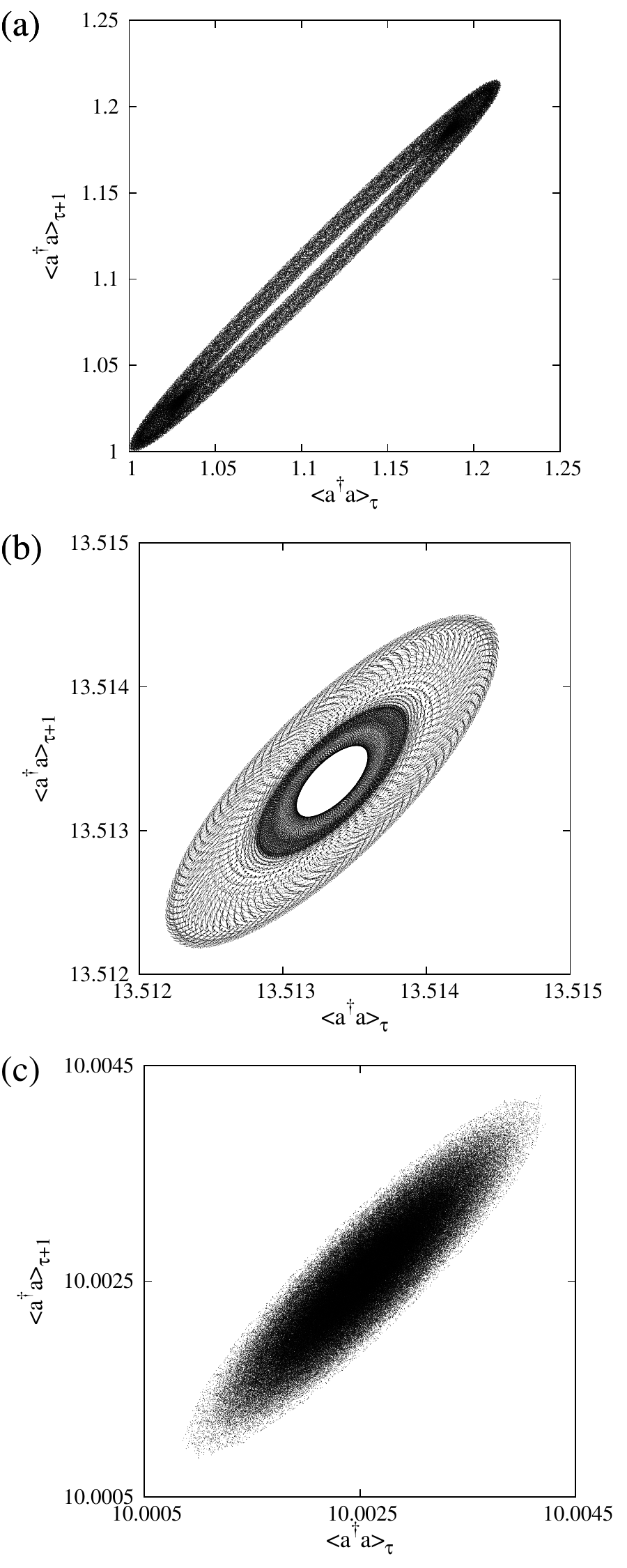}
\caption{Return maps of the mean photon number  $\langle
a^{\dagger}a
\rangle$ in the bipartite system. Initial states (a) $\ket{1; \alpha}$
${|\alpha|}^{2} = 1$, (b) $\ket{1; \alpha, 10}$
${|\alpha|}^{2} = 1$, and (c) $\ket{1; \alpha}$
${|\alpha|}^{2} = 10$}
\label{fig:single_mode_v_return}
\end{figure}

Turning to the tripartite (or two field modes plus atom) model,
we have studied extensively the 
recurrence statistics of the mean photon number 
of one of the field modes, 
for various initial states, and for 
different ratios of the strength of the nonlinearities to the corresponding 
interaction strengths. In order to compare our results with the bipartite model 
considered above,  and with the results of \cite{sudhrecurrence}, 
we present relevant plots 
in the illustrative case of the   
initial state $\ket{1; \alpha; \alpha}$ with ${|\alpha|}^{2} = 10$, 
setting $\chi_{1} / \lambda_{1} = \chi_{2} / \lambda_{2} =  5$.   
A long time series ($3 \times 10^{5}$
data points with time step equal to unity) 
of the mean photon number $\aver{{a_{1}}^{\dagger} a_{1}}$ was 
generated for implementimg the 
FNN algorithm and computing the Lyapunov exponent.  

\begin{figure}[H]
\centering
\includegraphics[width=\textwidth]{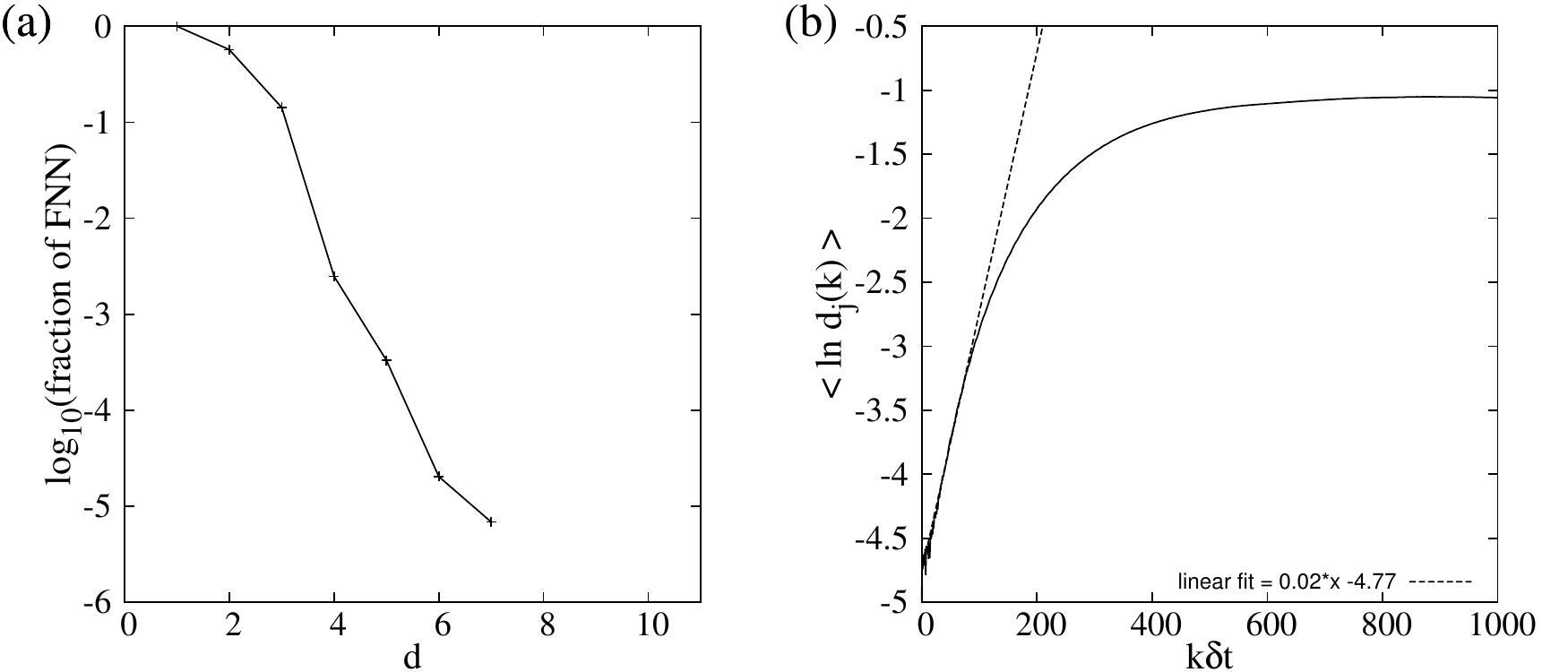}
\caption{(a) Fraction of FNN versus embedding dimension 
$d$, and (b) $\langle \ln d_{j} (k)\rangle$ versus $t (=k\delta t)$ using the 
algorithm of Rosenstein et al., for the initial state $\ket{1;\alpha;\alpha}$, 
${|\alpha|}^2 = 10$ for the tripartite system. The maximal Lyapunov exponent is 
$\approx 0.02$.} \label{fig:two_mode_v_fnn_ros}
\end{figure}
 
Figure 8(a) shows the fraction of FNN as a 
function of the embedding dimension $d$. 
We estimate the minimum embedding dimension to be $7$,
as the fraction of FNN becomes negligible 
($< 10^{-5}$) beyond this value. 
Using this, we have assessed the 
extent of the sensitivity to initial conditions displayed 
in the reconstructed phase 
space by computing the  
Lyapunov exponent from the time series of  $\aver{{a_{1}}^{\dagger} a_{1}}$.   
The initial set of separations $\{d_{j}(0)\}$ between the $j$th pair of nearest 
neighbours evolves to $\{d_{j}(k)\}$ after $k$ time steps. As is well 
known, the 
maximal Lyapunov 
exponent is the slope of the linear region that 
lies between the initial transient 
and the final saturation region in the plot 
of $\aver{\ln\, {d_{j}(k)}}$ versus time 
(figure 8(b)). 
Here, the average is over all values of $j$. The maximal Lyapunov 
exponent is estimated to be $\approx 0.02$, 
indicative of weakly chaotic behaviour. 
On the other hand,  
the corresponding first-return-time distribution 
to a generic initial cell (figure 9) obtained 
from a sample set of $10^{7}$ data points for 
the same initial state and parameter 
values is spiky, contrary to what we would expect.
We note, however, that the spiky distribution shows signs of tending
to an exponential one, 
as can be seen from the envelope of the spikes.
Once again, the low dimensionality of the atomic Hilbert 
space appears to be responsible for these features.

 \begin{figure}[H] \centering
\includegraphics[width=0.5\textwidth]{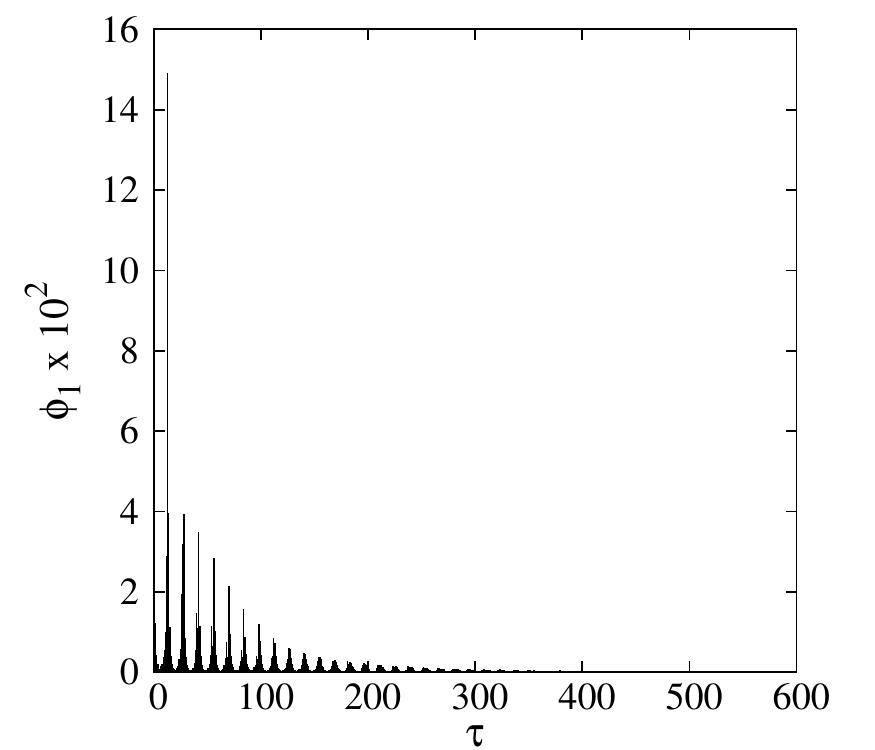}
\caption{ First-return-time distribution $\phi_{1}$ of the mean photon number
$\langle {a_{1}}^{\dagger}a_{1} \rangle$ in the
tripartite model. Initial state  $\ket{1; \alpha;\alpha}$ with
 ${|\alpha|}^{2} = 10$} 
\label{fig:two_mode_v_rtd}
\end{figure} 

\section{Concluding remarks}
Our investigations reveal interesting and somewhat counter-intuitive results on 
the dynamics of interacting atom-field systems. Independent of whether  
the radiation 
fields in a multipartite system  exhibit ideal coherence 
or significant departures from coherence, the number statistics and the 
extent of entanglement of the various subsystems 
with each other during temporal 
evolution are affected significantly by the `smallness' of the atomic Hilbert space. 

The recurrence time distribution $\phi_{1}$ of an observable to cells in a 
coarse-grained phase space of relevant observables of a system is an important 
quantifier of the nature of the dynamics. 
A spiky $\phi_{1}$ is generally indicative 
of quasiperiodicity, 
while an  
exponential distribution is a generic characteristic 
of hyperbolicity in the dynamics. The latter need not imply, however,
the positivity of any of 
the Lyapunov esponents.
Similarly, a spiky form of 
$\phi_{1}$ could accompany weakly chaotic
behaviour as indicated by a small positive Lyapunov exponent.
Both these uncommon features are exhibited 
by the models we have studied in this paper.

The strength of the nonlinearity in the field 
modes is the underlying feature that 
leads to a wide 
range of ergodicity properties of the mean photon number. 
In the presence of a sufficiently strong atom-field coupling, 
the field mode  evolves through a spectrum of 
states with different number statistics. 
It may be possible to reconstruct the 
latter through continuous variable quantum state tomography. 
An appropriate weak 
local probe can be used 
to determine the 
mean photon number of the field, and  
homodyne correlation 
measurements on the lines suggested in \cite{shchukin, webb} can perhaps be used to 
measure relevant observables.

\end{document}